\documentclass{elsarticle}

\newcommand{\omitfigs}{0}

\usepackage{amsmath}  
\usepackage{amsfonts}  
\usepackage{graphicx}  
\usepackage{epstopdf}  
\usepackage{amssymb}

\usepackage{epstopdf}

\usepackage{pgfplots}
\usetikzlibrary{pgfplots.groupplots}

\newcommand{\mytikz}[2]
{\ifodd \omitfigs
\else
\begin{tikzpicture}[scale=#1] \input{./figs/#2} \end{tikzpicture}
\fi
}

\newcommand{\be}{\begin{eqnarray}}
\newcommand{\ee}{\end{eqnarray}}

\newcommand{\kB}{k_{\rm B}}

\newcommand{\dby}[2]{ \frac{{\rm d} #1}{{\rm d} #2}}

\newcommand{\ket}[1]{ \left| #1 \right> }

\begin{document}
\def \d {\rm d}

\ifodd 1
Highlights

1. Momentum diffusion in an ordinary gas owing to Unruh radiation associated with the collisions is estimated.

2. The effect is sufficient to scramble the motion of the molecules on nanosecond timescales at standard pressure and temperature.

3. Effects such as this illustrate that quantum field theory implies that physical systems are open, not isolated, unless extreme measures are taken to modify the quantum vacuum.

4. It follows that the evolution of the observable universe is not unitary and thermodynamic irreversibility is a fact not an approximation.
\fi

\title{Irreversible behaviour of a gas owing to Unruh radiation}

\author{Andrew M. Steane}


\address{Department of Atomic and Laser Physics, Clarendon Laboratory, Parks Road, Oxford OX1 3PU, England.}

\date{\today}

\begin{abstract}
When gas molecules collide, they accelerate, and therefore encounter the 
Fulling-Davies-Unruh and Moore-DeWitt effects. The size of these effects is sufficient to randomize the motion of the gas molecules after about 1 nanosecond at standard temperature and pressure. Such observations show that quantum field theory modifies what is
required to isolate a physical system sufficiently for its behaviour to be 
unitary. In practice the requirements are never satisfied exactly. 
Therefore the evolution of the observable universe is non-unitary and thermodynamically irreversible.
\end{abstract}

\begin{keyword}
quantum chaos \sep Unruh effect \sep irreversibility \sep
quantum field theory \sep quantum foundations
\end{keyword}




\maketitle

In the first part of the argument we consider the exponential growth of
the separation of trajectories in a sequence of hard sphere collisions,
and of the out-of-time-order correlator
in a quantum treatment of a gas. In
the second part we consider momentum diffusion owing to 
effects associated with the quantum vacuum
in the case of bodies undergoing otherwise elastic collisions.
In the final part 
we combine the results so as to estimate the randomizing effect of its own Unruh 
radiation, and Moore-DeWitt radiation, 
on a gas which is otherwise isolated. Implications for irreversibility 
and determinism in general are discussed.

\begin{figure}
\centering
\begin{tikzpicture}
\begin{scope}[rotate=0,scale=0.7]
    \draw (0,0) circle (1.5);
	\draw [thin] (0,-1.5)--(0,0)--(0.7,-1.33);
	\draw [thin] (0,0)--(0.83,-1.26);
	
	\draw (0.2,-0.9) node {$\theta$};
	
	\draw [->, >=latex] (0.7,-2.8)--(0.7,-1.33)--(2.5,-2.4);			
	\draw [->, >=latex] (0.83,-2.8)--(0.83,-1.26)--(2.8,-2);
\end{scope}		
\end{tikzpicture}
\caption{The final direction of motion when one sphere collides with another
changes by $2 \delta \theta$ when the angle $\theta$ changes by $\delta\theta$.}
\label{f.sphereangle}
\end{figure}
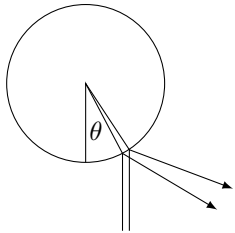

Consider a collection of rigid spheres in motion and colliding with one another 
in elastic collisions, as described by classical physics. This is a simple model 
of a gas so we will refer to these spheres as `atoms'.  
Suppose that the direction of motion of some given atom 
is uncertain by $\Delta \theta_0$. If the distance travelled by an atom
between collisions is $\lambda$, then the
position at which two atoms hit one another is imprecise by
approximately $\lambda \Delta \theta_0$. If the atomic radius is $r$ then the collision
point subtends an angle at either atom's centre that is uncertain by approximately
$\lambda \Delta\theta_0/ r$. The collision causes the direction of motion
after the collision to become uncertain by an amount of order twice this
(see figure \ref{f.sphereangle}), so the direction of motion after
the collision is uncertain by $\Delta\theta_1 \simeq 2 (\lambda/r) \Delta \theta_0$. 
By the same argument, the direction of motion after two collisions is uncertain by
$
\Delta \theta_2 \simeq 2 (\lambda/r) \Delta \theta_1 = 4 (\lambda/r)^2 \Delta\theta_0.
$
Further collisions continue to amplify the uncertainty. After $n$ collisions, and
assuming the angles remain small compared to one radian, the uncertainty
is of order
\be
\Delta \theta_n \simeq (2 \lambda/r)^n \Delta\theta_0.
\ee
The exponential growth here is an example of a Lyapunov exponent. Our simple
calculation matches quite well the results of
more sophisticated treatments in the
literature.\cite{Latz1997}
For the quantum treatment one may study the
out-of-time-order correlator which is closely related to the classical 
Lyapunov exponent.\cite{Larkin1969,Maldacena2016} 
Zhang obtains values consistent with the above for a gas at high
temperature (far from degeneracy).\cite{Zhang2019}.
On this basis we claim that, at high temperatures,
the number of collisions required to randomize
the motion of a
gas is the number required
to obtain $\Delta\theta_n \sim 1$:
\be
n \simeq - \frac{ \ln \Delta\theta_0 }{\ln (2 \lambda/r)}.
\label{nrandomize}
\ee
This concludes the first part of the argument.

Now let's consider Unruh radiation \cite{76Unruh,73Fulling,08Crispino}. 
The Fulling-Davies-Unruh effect can be presented in various
ways \cite{08Crispino,84Unruh,05Fulling,80Boyer,Scully2021,Scully2022}. 
(A related phenomenon is the Moore-DeWitt effect, also called 
(by a misnomer) the dynamical Casimir effect \cite{Moore1970,DeWitt1975,Paraoanu2020,00Dalvit,Dodonov2009}.)
At the heart of it is Unruh's observation that, in the coordinate system of the Rindler frame 
(constantly accelerating frame in flat spacetime), the vacuum state of quantum field theory takes the form of a thermal state with temperature
$
T_{\rm U} = 
{\hbar a} / {(2\pi \kB c)} 
$
where $a$ is the proper acceleration.
There is not yet a consensus on the detail of the physical 
implications \cite{04Fulling,05Fulling,Grove1986,06DeBievre}, but it will be sufficient to our purpose
to take the following broadly standard point of view. We consider an {\em inertial} reference
frame, with the electromagnetic field initially in its vacuum state.
In this frame we suppose there exists a system A which provides a force on system B,
causing B to accelerate. B may be any system having electromagnetic interactions,
such as a charged body, a detector, an atom, a lump of fused silica. In this situation, we claim, the net force on B will fluctuate. 
If B has internal structure, then
it will undergo internal excitations, and subsequently emit photons, in a stochastic way.
The energy is provided by system A; the complete process
involves the system's response to correlations in the vacuum state of
the electromagnetic field.\cite{Scully2021,Scully2022}
If the proper acceleration is constant on average 
and goes on for long enough, then
the resulting fluctuation is the same as if B were bathed in thermal radiation 
at the Unruh temperature in its instantaneous rest frame. 
If B has no internal structure but is an accelerating
charged particle, the QED treatment of radiation reaction equally 
leads to a fluctuating force \cite{91Ford,05Fulling,Kolekar2012,Kolekar2020}.

In view of the above, we make the following claim. A physical entity
accelerating through the vacuum will undergo internal
excitations similar to those it would undergo if it were moving inertially and
subject to radiation such that the density matrix of the electromagnetic field is diagonal in
the Fock basis, with a mean excitation per mode
\be
\bar{n}(\omega) = \left[ {\exp({2\pi c\omega/a})-1} \right]^{-1}.  \label{nbar}
\ee
A major source of imprecision in the following calculation 
is the fact that the motion under consideration
will only involve acceleration for short periods of time, so the Unruh result does not apply exactly. 
(It is also true that the acceleration during a particle collision is not constant, but we expect a
brief period of constant acceleration to be a reasonable approximation for rough calculations.)
Acceleration for finite periods is discussed in \cite{03Martinetti,08Crispino,08Barton}. The
approximation that the Unruh temperature applies to the majority of the elapsed
proper time is good when the product of proper acceleration and
proper time is of order $c$, but the motion we shall consider is far from that limit. 
To further support the main argument of the paper, 
we shall also present a rough
estimate from the Moore-DeWitt effect, after the Unruh-effect estimate. 

Momentum diffusion owing to scattering of thermal radiation is described by
Joos \cite{85Joos}. Adopting that model, 
we find that an accelerating body experiences momentum diffusion
such that after a time $\tau$
the momentum variance owing to Unruh radiation 
at frequency $ck$ is $\Delta p_k^2 = \hbar^2 k^2 \Gamma_k \tau$,
where 
\be
\Gamma_k = \int \dby{\sigma}{\Omega} {\rm d}\Omega \frac{\bar{n}c}{V} 2(1-\cos \theta)
\, \simeq 
 \frac{\bar{n}c}{V} \sigma
 \label{Gammak}
\ee
where $\sigma$ is the cross-section for scattering of radiation by the body,
$\bar{n}$ is given by Eqn (\ref{nbar}) with $\omega=ck$ and
$V$ is the volume of space containing the electromagnetic field; it will go to
infinity when we integrate over $k$.

It will emerge that the wavelengths concerned are in the radio wave part of the spectrum, where a suitable approximation for atoms and molecules is to
use the polarizability and hence scattering cross-section of a conducting sphere:
$
\sigma \simeq (10 \pi/3) r^2 (k r)^4.
$
We thus find for the total momentum diffusion
\be
\Delta p^2 = \int \Delta p_k^2  \frac{V \d^3 k}{(2\pi)^3} \, \simeq \,
\frac{5 \hbar^2}{3\pi r^2} \frac{c}{v} 
\int \frac{x^8}{\exp(\alpha x) - 1} \d x        \label{Dp2}
\ee
where $x = kr$ and $\alpha = 2 \pi c^2 / a r$. For two molecules of size
$r$ the acceleration during an elastic collision at initial speed $v$
can be estimated as $a \simeq v/\tau \simeq v^2/r$. For a gas at temperature
$T$, $\langle v^2 \rangle = 3 \kB T/m$ so we find $a \simeq 3 \kB T/ mr$
and $\alpha \simeq (2/3) \pi m c^2 / \kB T$. We shall be considering a gas far from the relativistic limit, so $\alpha \gg 1$. In this situation the integrand in (\ref{Dp2}) is sharply peaked at $\alpha x \simeq 8$ and therefore the $-1$ in the denominator can be neglected and the integral is well approximated
by $8! / \alpha^9$. 

Owing to the momentum diffusion associated with a single collision, the direction of motion of either molecule is liable to vary by the
standard deviation $\Delta \theta_0 = \Delta p / p$. Hence we find
\be
\Delta \theta_0 \simeq \frac{\hbar}{r m v} 
\sqrt{ \frac{8! 5 c}{3 \pi v}  }
\left( \frac{3\kB T}{2\pi m c^2} \right)^{9/2} .     \label{Dtheta}
\ee

We now apply the above to nitrogen gas at standard temperature and pressure.
We find $r m \bar{v} \simeq 103 \hbar$, $\alpha = 2\times 10^{12}$, so $\alpha k r = 8$ when $k = 0.007\,{\rm m}^{-1}$ which corresponds to a wavelength around 850 m. Eqn (\ref{Dtheta}) gives $\Delta \theta_0 \sim 3 \times 10^{-53}$ and then (\ref{nrandomize}) gives $n \simeq 34$. It takes about a nanosecond for any one N$_2$ molecule to undergo 34 collisions at STP. Therefore we find that nitrogen gas is randomized by the effects of Unruh radiation (amplified by collisions) after about one nanosecond at STP. 

Now let's consider the Moore-DeWitt effect. We will adopt the results of Dalvit and Neto,\cite{00Dalvit} with appropriate adjustments. Dalvit and Neto (DN) consider a mirror of imperfect reflectivity oscillating in the direction
normal to its surface, coupled to a scalar quantum field. They use a coupling parameter $\Omega$ which is defined such that the amplitude
reflectivity is $R(\omega) = -i\Omega/(\omega + i \Omega)$. We will crudely
transfer this model to our situation by treating a single collision as an
oscillation of a mechanical oscillator through one half-period, with
frequency $\omega_0 = v/r$. The colliding atoms constitute the `mirror' and
we estimate the coupling parameter by adopting $k^2 \sigma \sim |R(\omega)|^2$
using the scattering cross-section $\sigma$ as above.  This gives
$\Omega \simeq \sqrt{10 \pi/3} (kr)^3 \omega$. DN find momentum diffusion at the
rate $\Gamma m \omega_0 \hbar/2$ where $\Gamma$ is given in their eqn (9) as
\be
\Gamma = \frac{\hbar \Omega \omega_0}{2 \pi m c^2} \zeta(\omega_0/\Omega)
\simeq \frac{\hbar \Omega^2}{2 \pi m c^2} {\ln( \omega_0/\Omega)},
\ee
where $\zeta(u) = \ln(1 +u^2)/2u - \arctan(u)/u^2 - 1/u$ and the second
expression is for the case $\Omega \ll \omega_0$. Using $\Delta p^2 = \Gamma m \omega_0 \hbar \tau/2$ we obtain
\be
\Delta \theta_0 = \frac{\Delta p}{p} \simeq \frac{\hbar \Omega}{2 m v c}
\sqrt{ \ln(\omega_0/\Omega) }
\ee
where we used $\omega_0 \tau = \pi$. Our expression for $\Omega$ gives it as
proportional to $\omega^4$. We must use the value $\omega \sim \omega_0$ in
this expression because that is the order of magnitude of the range of 
frequencies sampled by the (one half-period) `oscillation'. This gives
$kr = v/c$ and we obtain
\be
\Delta \theta_0 \simeq \frac{\hbar}{r mv} \sqrt{(5\pi/6) \ln(\omega_0/\Omega)}
\left( \frac{v}{c} \right)^4 .   \label{DthetaM}
\ee
Ignoring for a moment the $\sqrt{\ln}$ term in (\ref{DthetaM})
(which has only a weak dependence on $v/c$), 
(\ref{Dtheta}) and (\ref{DthetaM}) both have the form $\Delta \theta_0 \propto (\hbar/rmv) (v/c)^n$ but with different powers $n$. 
Even if we have estimated the coupling $\Omega$ poorly, the result for $\ln \Delta \theta_0$ will not change by much. 
We find that the Moore-DeWitt effect dominates the Unruh
effect in the collisions under discussion, and the
value of $\ln(\Delta \theta_0)$ is predicted to be about twice as large as the one predicted from the Unruh effect, so 16 collisions are sufficient to randomize the motion at STP.
It is not necessary for either calculation to be very accurate in order
that they are sufficient to support the main argument
of this paper. 

There are many larger physical effects which will influence the motion of
gas molecules more quickly than this. We have drawn attention to
the role of quantum field theory (QFT) because of the implications for determinism and
reversibility in physics more generally. The Fulling-Davies-Unruh effect,
Hawking radiation and the Moore-DeWitt effect all call into question the degree to
which QFT predicts deterministic, unitary evolution.
One can observe from the role of Schr\"odinger's equation in QFT that it predicts globally unitary behaviour of an entirely isolated system. However 
QFT modifies our understanding of what is required to achieve such
isolation. It is not sufficient merely to put other systems far away. One
must also `switch off' the vacuum effects by banishing black holes and
surrounding the system in question with perfectly reflecting walls. 

Both the Unruh and Hawking calculation (which
are related) can be formulated by expressing the vacuum state 
in the entangled form
\be
\ket{0} &=& \prod_i  C_i \sum_{n_i=0}^\infty e^{-\pi n_i \omega_i c/a} \ket{n_i,R} \otimes \ket{n_i,L}
\ee
where $C_i$ are constant coefficients and $R,L$ refer to spacetime
regions which may be spacelike-separated and on different sides of
a horizon. In order to describe physical
effects in one region one traces over the variables associated with the other.
Taking the trace corresponds to throwing away information and this is 
what leads to non-unitary evolution. The evolution of the density matrix
remains deterministic, but the realisation of physical outcomes is not entirely
deterministic because the physical system adopts one state or another
from a pointer basis---the prediction for a Schr\"odinger cat, for
example, gives probabilities of one half 
but in any given trial the cat is either alive or dead. (A more careful
statement would be: a rational observer wishing to plan for their future is
informed by quantum theory that one or the other outcome, not evidence of
a superposition of both, will be observable, and the theory furnishes the
probabilities and no more.)

It is well known that open quantum
systems can exhibit non-unitary behaviour which can be modelled by
calculating their entanglement with other systems and then tracing over the
latter.\cite{Rivas2012} A lesson from effects associated
with the QFT vacuum is that all physical systems are open in this sense,
and therefore liable to have non-unitary dynamics, unless measures
are taken to modify the QFT vacuum by the use of mirrors of extremely high 
reflectivity. In short, QFT does not predict unitary behaviour of any part of 
the universe, nor even of the whole observable universe, except in
certain idealized circumstances which never hold exactly. This same
conclusion was affirmed by Unruh and Wald in quite general terms.\cite{Unruh2017} 
They discussed it more fully in the context of black holes; we have
shown that such considerations affect everyday phenomena.

A related issue is what kind of physical system warrants comparison
with an ordinary household cat, when one is considering issues in
quantum interference and measurement raised by the Schr\"odinger's cat
thought-experiment. Among the significant features of a cat is the chaotic motion of its water molecules, such that they exhibit the kind of sensitivity
to their own Unruh radiation which we have discussed. No physical system 
exhibiting empirical evidence of overall quantum superposition has ever been even
remotely comparable to a cat on this measure.

It would be of interest to calculate the QFT effects during a collision
more carefully than was done here. One must account, in particular, for the
very small duration of the periods of acceleration, where motion never approaches the speed of light, in the Unruh
effect. One might also wish to set up the Moore-DeWitt calculation afresh.
Such efforts may considerably modify the estimate of $\Delta \theta_0$. 
However if it does not change
$\log( \Delta\theta_0)$ by very much then we would still have the prediction of randomization of a gas on nanosecond timescales. 

I thank a thoughtful referee for drawing my attention to some further work and
for the insight that it would be valuable to flesh out the Moore-DeWitt
part of the argument more fully. 

\bibliographystyle{plain}

\end{document}